\documentclass[fleqn,10pt]{wlscirep}
\usepackage[utf8]{inputenc}
\usepackage[T1]{fontenc}
\usepackage{lineno}
\title{A Cross Spatio-Temporal Pathology-based Lung Nodule Dataset}

\author[1,*]{Muwei Jian}
\author[1,$\dag$]{Haoran Zhang}
\author[2,$\dag$]{Mingju Shao}
\author[3]{Hongyu Chen}
\author[1]{Huihui Huang}
\author[1]{Yanjie Zhong}
\author[4]{Changlei Zhang}
\author[4]{Bin Wang}
\author[2]{Penghui Gao}

\affil[1]{School of Computer Science and Technology, Shandong University of Finance and Economics, Jinan, China}
\affil[2]{Emergency Medicine Center, The Second Hospital, Cheeloo College of Medicine, Shandong University, Jinan, China}
\affil[3]{School of Information Science and Technology, Linyi University, Linyi, China}
\affil[4]{Medical Imaging Center, The Second Hospital, Cheeloo College of Medicine, Shandong University, Jinan, China}

\affil[*]{corresponding author(s): Muwei Jian (jianmuwei@ouc.edu.cn)}

\affil[$\dag$]{They contribute equally to this work and share the first authorship.}

\begin{abstract}

Recently, intelligent analysis of lung nodules with the assistant of computer aided detection (CAD) techniques can improve the accuracy rate of lung cancer diagnosis. However, existing CAD systems and pulmonary datasets mainly focus on Computed Tomography (CT) images from one single period, while ignoring the cross spatio-temporal features associated with the progression of nodules contained in imaging data from various captured periods of lung cancer. If the evolution patterns of nodules across various periods in the patients’ CT sequences can be explored, it will play a crucial role in guiding the precise screening identification of lung cancer. Therefore, a cross spatio-temporal lung nodule dataset with pathological information for nodule identification and diagnosis is constructed, which contains 328 CT sequences and 362 annotated nodules from 109 patients. This comprehensive database is intended to drive research in the field of CAD towards more practical and robust methods, and also contribute to the further exploration of precision medicine related field. To ensure patient confidentiality, we have removed sensitive information from the dataset.

\end{abstract}
\begin{document}
\captionsetup[figure]{name={Fig.}}

\flushbottom
\maketitle

\section*{Background \& Summary}

Effective lung cancer screening and diagnosis has emerged as the critical factor in prolonging patient survival. Among the spectrum of fatal diseases, pulmonary diseases constitute a significant portion \cite{leiter2023global, weiss2023deep, fu2023gender}. Within this category, lung cancer stands out as the most prominent, and its incidence and mortality rates are growing the fastest. This renders it one of the formidable malignancies, posing a severe threat to the population’s health and well-being \cite{siegel2023cancer, badr2023unified}. Early diagnosis and prompt treatment of lung cancer patients have been shown to substantially enhance their survival rate \cite{rudin2021small, howlader2020effect, prior2017public}. Since pulmonary nodules as the initial clinical manifestation of lung cancer, the accurate diagnosis and thorough screening of pulmonary nodules assume pivotal roles in the overall therapy of lung cancer \cite{venkadesh2023prior, mazzone2022evaluating, kim2022artificial}. Medical imaging techniques encompassing modalities such as X-ray \cite{reis2022brax}, CT \cite{wang2022follow, li2022cov}, Positron Emission Tomography (PET), and so on, take on a critical function in the screening process for lung cancer. Among these techniques, CT scanning is regarded as a routine and indispensable tool in clinical practice, capable of non-invasively capturing the intricate heterogeneity of lung tumours. Consequently, it has emerged as the primary modality for lung cancer detection in contemporary healthcare settings.

The conventional diagnostic approach primarily involves clinicians manually examining CT images slice by slice and relying on their clinical expertise to make a diagnosis. However, this approach has several limitations \cite{chen2021annotation}. Firstly, radiologists review hundreds of lung CT scans daily, which is a time-consuming and labour-intensive task. Secondly, diagnostic decisions made by clinicians can be highly subjective and may be influenced by their individual levels of expertise and clinical experience. Moreover, the heterogeneity of malignant lung nodules is not solely manifested in the size and morphological characteristics of the lesions, but also in subtle variations in local pixel values within CT images. These trivial nuances are often challenging for the human eye to discern accurately. As a result, accurately determining the malignancy degree of nodules poses a formidable challenge for physicians \cite{paez2023longitudinal}.

The implementation of CAD systems in the context of lung nodule detection effectively addresses the aforementioned challenges \cite{liang2016low, papanastasiou2023focus}. The traditional methods include those based on morphological features, voxel-based clustering, and threshold-based techniques \cite{de2014automatic}, which are highly reliant on manual hand-crafted feature representation and extraction. In recent years, the advent of deep learning-based algorithms applied to medical images \cite{shin2012stacked, paez2023longitudinal}, and has led researchers to progressively propose 2D convolutional neural networks \cite{wang2019higher, cheng2016computer, xie2018fusing} and 3D convolutional neural networks \cite{liao2019evaluate, zhou2019models, zhu2018deeplung, li2020deepseed, zhou2023unified} for the purpose of diagnosing lung nodules. These advancements in lung cancer CAD systems have resulted in breakthroughs and remarkable developments, yielding more accurate results than those traditional methodologies.

At present, the majority of lung nodule detection research relies on the LIDC-IDRI dataset \cite{armato2011lung} or its extended subsets. However, they lack validation through pathology reports, leading to some degree of uncertainty regarding the authenticity of the labels. Additionally, while some new datasets have emerged in recent studies \cite{mei2021sanet, shao2022lidp}, they all share a common and distinct limitation - covering only clinical imaging data at a single time point, thus overlooking the crucial properties, trends, and patterns associated with the gradual evolution of tumors cross spatio-temporal in different periods.
	
The pattern variation of nodules is constantly evolving and morphing over various times. In the bargain, it is noteworthy that patients often visit distinct healthcare facilities and undergo multiple CT scans over the course of their illness at different periods. Nevertheless, challenges in the interoperability and seamless exchange of medical information persist, hindering the effective sharing of medical data among individual hospitals at various times. In clinical practice, data from different periods and sources is beneficial for obtaining a comprehensive judgment, however, physicians realistically encounter difficulties in monitoring and access to medical information associated with the patient’s CT scans from diverse periods. This isolation of information can impact the consistency of a physician’s overall diagnosis analysis and treatment plan, detrimentally affecting patient therapy potentially. In light of this, to be viewed as a dynamic observational and evolutionary perspective will be instrumental in tackling the task of lung nodule detection in CAD system. This process relies on the utilization of cross spatio-temporal series data of different periods to identify and analyze the regularities of the disease across times according to the variation of its development. Briefly, establishing the disease evolution model and enabling the provision of disease dynamic insights for physicians will be conducive to generating diagnostic conclusions as well as forecasting the future disease state for the patient. Therefore, a diversified dataset with both temporal and spatial dimensions across various times is urgently needed.

In this study, we introduce a novel cross spatio-temporal lung nodule dataset based on pathological information, which effectively integrates rich multimodal information within the spatio-temporal dimension. The constructed dataset comprises 328 CT scans from 109 patients, including 250 slices with malignant nodules and 112 slices with benign nodules. Moreover, the dataset includes 182 cross spatio-temporal CT scans of different times. During the processing of the dataset, we accurately locate and annotate the nodules in the CT scans according to the pathological information. This critical step ensures the precise tracking and quantitative analysis of nodules, thereby furnishing reliable clinical data (e.g., Ground truth with labeled detail) for disease research and medical imaging analysis. To the best of our knowledge, this dataset is the pioneering pulmonary nodule dataset that embraces the concept of cross spatio-temporal integration with various periods, and provides a fresh dynamic perspective and opportunity for future medical imaging investigations.

\section*{Methods}

In order to more effectively address the clinical need for mining richer information related to lung disease, discovering potential imaging features, and predicting the evolutionary tendency of the disease, we partner with The Second Hospital of Shandong University to construct a pathology-driven cross spatio-temporal lung CT dataset drawn from actual clinical cases. This dataset is correct annotated based on pathology information. The data samples involved in this study have been reviewed and supported by the research ethics committee of the Second Hospital of Shandong university (approval number: KYLL-2023LW089), since this dataset were collected retrospectively, the ethics committee has waived the requirement for researchers to provide informed consent from patients. The following section will provide a detailed overview of the dataset, covering aspects such as the collection principles, data collection and annotation, dataset structure, dataset properties and so on.

\subsection*{Collection principles}

To ensure the accuracy of the collected subjects and to eliminate interference from irrelevant factors, we implement the following principles in our case selection process:

\textbf{Inclusion Criteria:} In accordance with pathology-driven criterion, only pulmonary nodules with a definitive pathological diagnosis are included. These nodules must be uniquely identifiable in the corresponding CT images based on the provided descriptions of nodule location and size in the diagnosis. This strict criterion ensures that each target nodule aligns precisely with its pathological diagnosis.

\textbf{Exclusion Criteria:} Cases with pathological diagnoses unrelated to pulmonary nodules are excluded. This step effectively eliminates any interference from other non-nodule lesions in our dataset.

\textbf{Pre-Treatment Imaging:} All CT scans are performed before any relevant treatments are administered. This inclusion criterion ensures that the image characteristics of the nodules are not influenced by prior treatments.

\textbf{Quality Control:} We meticulously review and discard CT images with missing layers or incorrect layering to guarantee the integrity and continuity of the lung nodule image dataset.

\subsection*{Data Collection and Annotation} 
Our data collection and annotation efforts are conducted at The Second Hospital of Shandong University. The data collection process encompasses four key stages, as illustrated in Fig. \ref{fig:Fig. 1}:

\begin{figure}[ht]
\centering
\includegraphics[width=\linewidth]{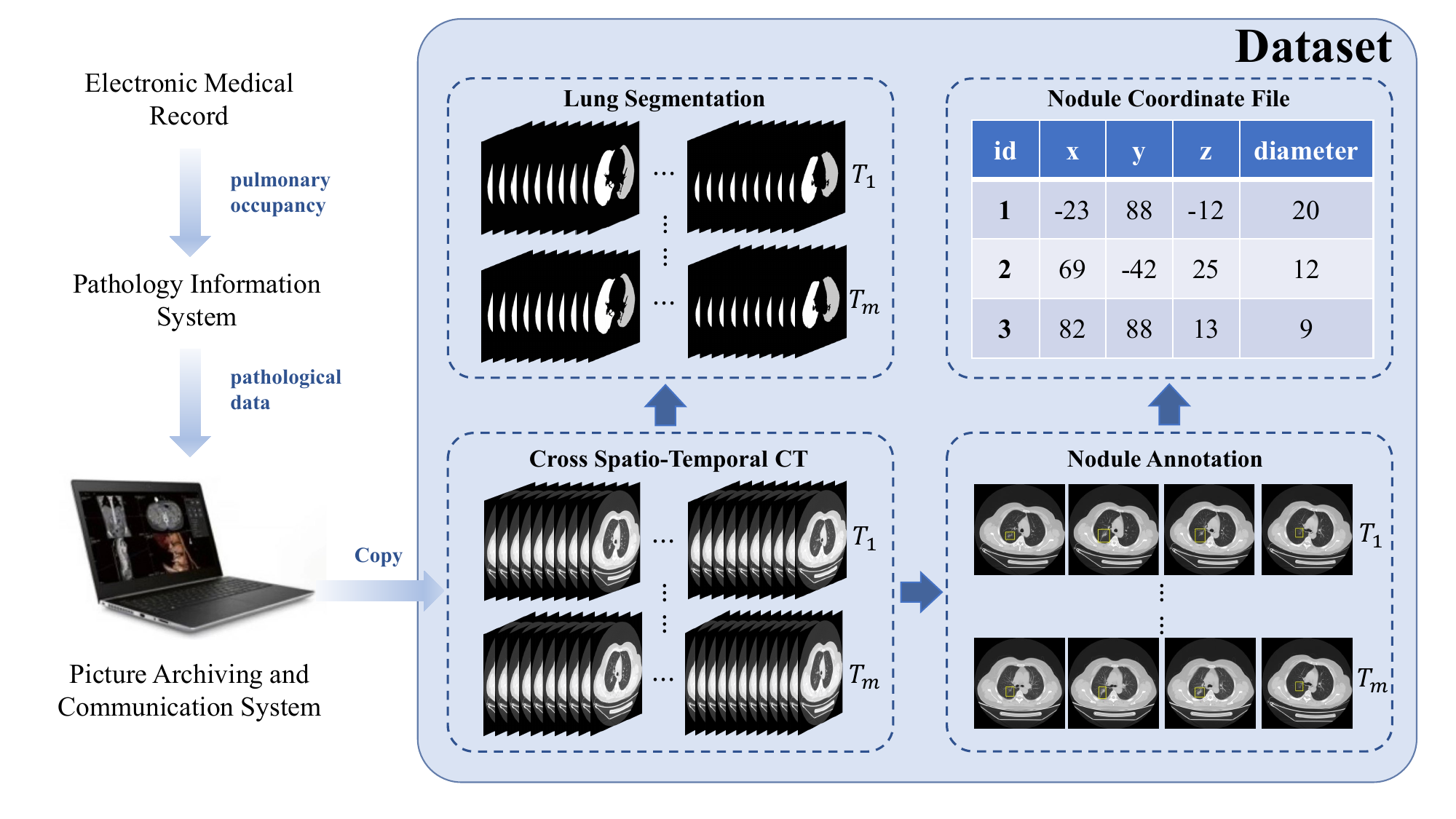}
\caption{The workflow for generating the dataset. Firstly, the Electronic Medical Record System 
(EMRS) is used to identify cases diagnosed with pulmonary occupying lesions within the past six years. Subsequently, these cases are filtered using the Pathology Information System (P.I.S), retaining only those with available pathology information. Finally, one or multi-time CT sequences of the patient are exported from the Picture Archiving and Communication System (PACS). After data extraction, the dataset is categorized into two distinct categories: classification and detection. The expert doctor extracts the coordinate information of the nodules from the CT sequences, based on the corresponding pathological information, and records it in the CSV file. In addition, post-segmentation lung data is also provided.
}
\label{fig:Fig. 1}
\end{figure}

\textbf{Case collection:} We amassed cases from the EMRS spanning from January 2016 to January 2023, specifically selecting those cases with diagnoses referencing nodular lung lesions and accompanying surgery records.

\textbf{Pathological diagnosis recording:} On the basis of the organized case records, we retrieve and document the pathological diagnosis corresponding to the lung nodules in the P.I.S.

\textbf{Patient imaging data retrieval:} In the PACS system, we search for the CT scan times of the patients, view the patients’ multiple diagnostic imaging reports and CT images, and subsequently employ the provided image descriptions in the diagnostic imaging reports along with the corresponding pathological diagnostic results to precisely locate the target pulmonary nodule.

\textbf{Data export and archive preparation:} Following the identification of the target pulmonary nodules, we proceed to export the corresponding DICOM file sequences. To facilitate subsequent labelling operations and ensure data preservation, we create archive disks containing the imaging data.

Under the guidance of expert physicians, we perform the labelling of nodule location and contour within the exported CT sequences. Then physicians verify the lesion locations by referencing the imaging manifestations and the pathological information contained in the dataset. Afterwards, they instruct the annotators to complete the marking/labelling task within the slices containing the lung nodules. This rigorous process is crucial in guaranteeing the precision and reliability of the annotations, forming a robust foundation for subsequent data analysis. In total, we label 362 nodules across 328 CT scans in this dataset.

\subsection*{Dataset Structure}

\begin{figure}[ht]
\centering
\includegraphics[width=\linewidth]{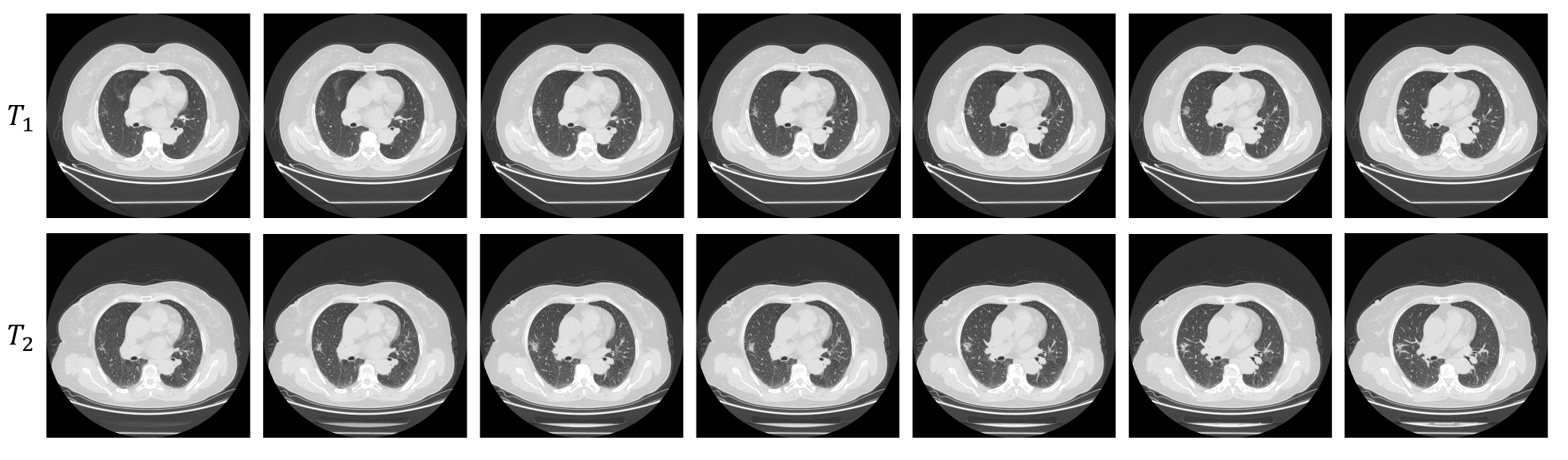}
\caption{An example of the CT sequences of different temporal dimension.
}
\label{fig:Fig. 2}
\end{figure}

\textbf{Temporal dimension CT series:} We collect CT scans from patients at various time points, denoted as $T_{1}^{},T_{2}^{},\cdots,T_{m}^{}$
, where $m$ represents the CT scans captured at the $m$ moments for each patient. Unlike single-point-in-time slice studies, longitudinal studies yield dynamic data tend to unearth richer information about the disease variation. Fig. \ref{fig:Fig. 2} shows an example of CT images taken at different times for an identical patient.

\begin{figure}[ht]
\centering
\includegraphics[width=\linewidth]{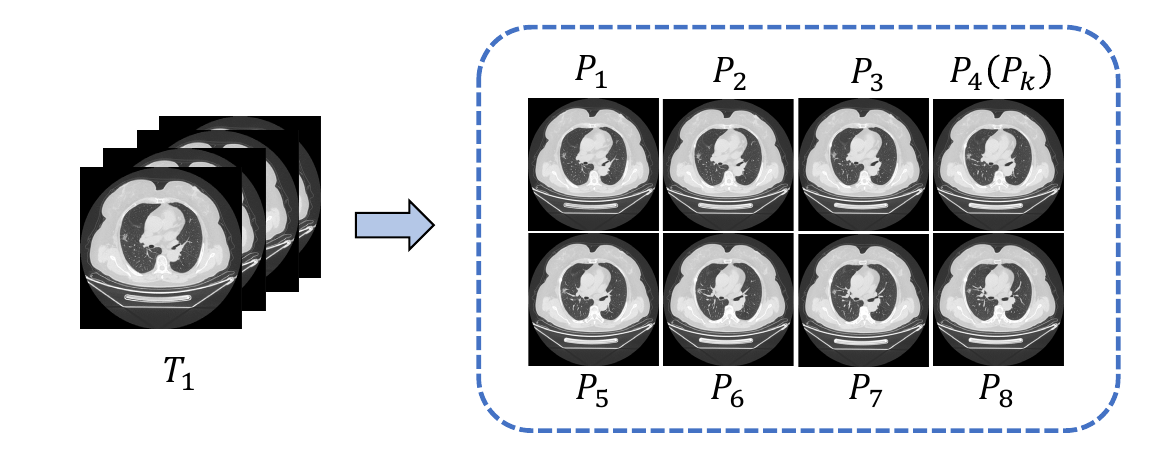}
\caption{An example of the CT sequences in terms of spatial dimension.
}
\label{fig:Fig. 3}
\end{figure}

\textbf{Spatial dimension CT Series:} In the spatial dimension, we annotate the CT scan sequences (referred to as $ P_{1}^{}\!,\!P_{2}^{}\!,\!\cdots\!,\!P_{k}^{}\!,\cdots\!,\!P_{n}^{}$) containing pulmonary nodules. Within these sequences, we identify and designate specific slices where the nodules are most prominently visible as keyframes, represented by $P_{k}^{}$. Fig. \ref{fig:Fig. 3} provides an example of the CT sequences for a patient in line of the spatial dimension.

\begin{figure}[ht]
\centering
\includegraphics[width=\linewidth]{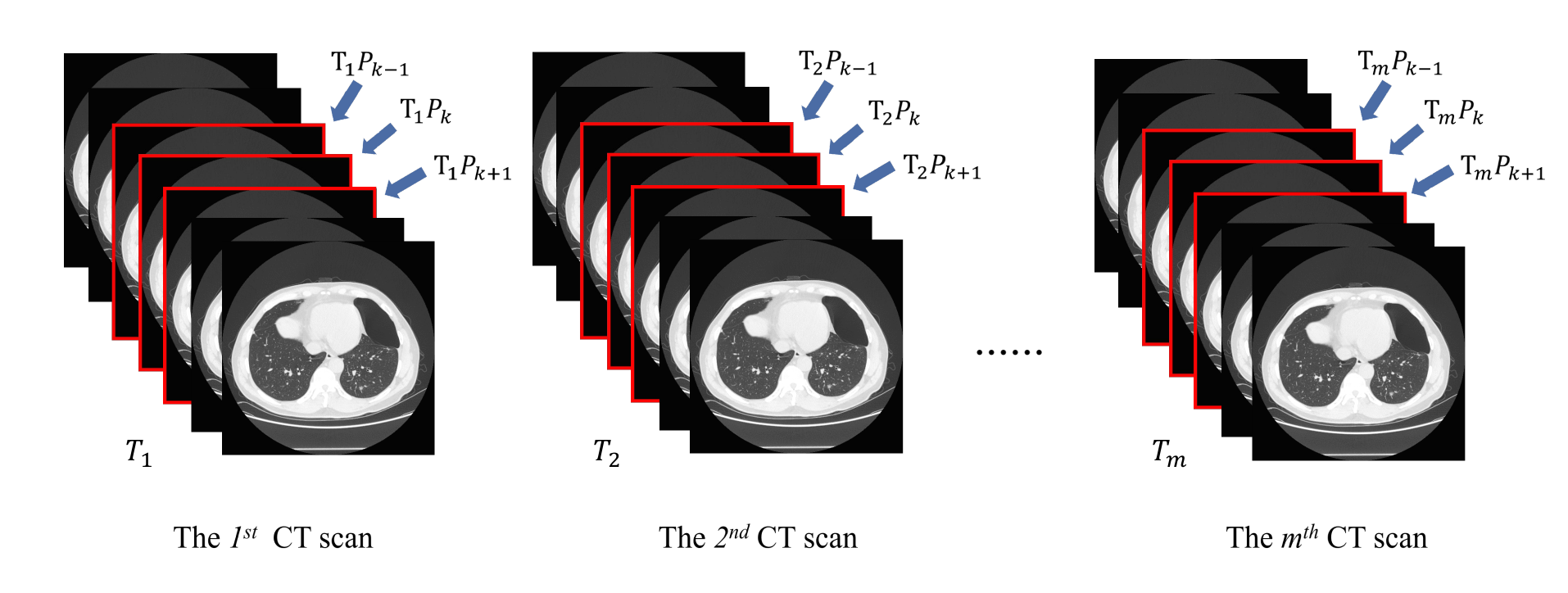}
\caption{A global overview of the dataset structure.
}
\label{fig:Fig. 4}
\end{figure}

\textbf{Cross spatio-temporal dimension CT Series:} Our dataset encompasses both longitudinal inspection data spanning multiple time moments and spatially oriented horizontal data. This comprehensive dataset enables researchers to concurrently observe nodule changes from the temporal and spatial perspective simultaneously, facilitating the training of accurate lung nodule detection models and providing a rich, multi-dimensional perspective for in-depth exploration of disease progression. A holistic overview of the dataset is presented in Fig. \ref{fig:Fig. 4}.

\subsection*{Dataset Properties}

\begin{figure}[ht]
\centering
\includegraphics[width=\linewidth]{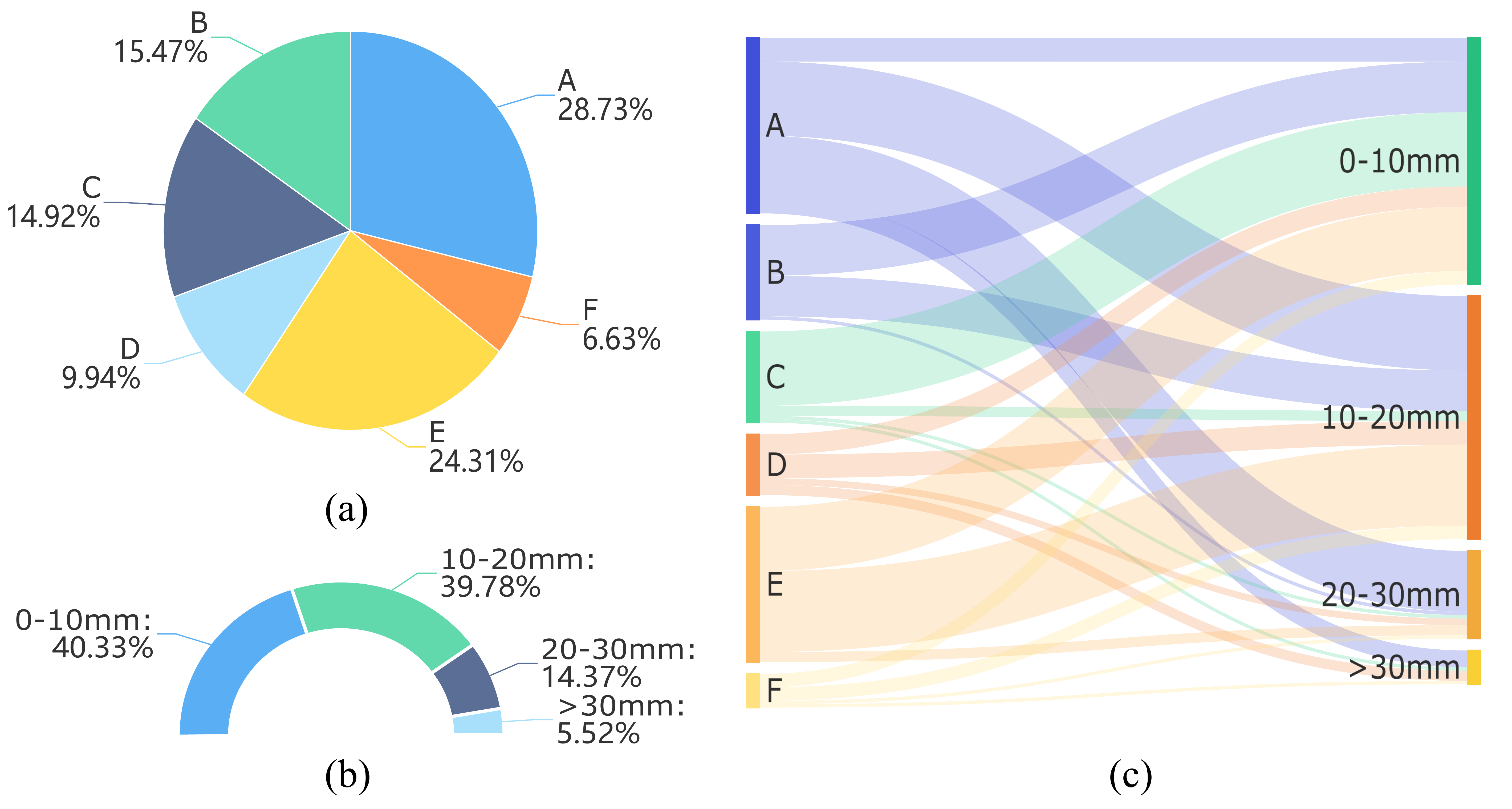}
\caption{Dataset properties. (a) The type and percentage of pathology in the dataset; (b) The type and percentage of nodule size; (c) Mutual dependencies among nodule size and pathologic type.
}
\label{fig:Fig. 5}
\end{figure}

\textbf{Pathological type:} Since the CT data in this dataset originate from patients who have underwent surgery or medical treatment, there are more malignant cases than benign cases. To address this imbalance while also considering clinical relevance, we applied a like-for-like aggregation approach to classify the dataset into six distinct pathology types, encompassing invasive adenocarcinoma (A), minimally invasive adenocarcinoma (B), adenocarcinoma in situ (C), malignant subtypes other than the aforementioned (D), cases involving inflammation (E), and benign subtypes (F), as illustrated in Fig. \ref{fig:Fig. 5}(a).

\textbf{Nodule Size:} Pulmonary nodules have been categorized based on their size into four distinct groups: 0-10 mm, 10-30 mm, 20-30 mm, and >30 mm. Among these size-based categories, pulmonary nodules within the range of 0 to 10 mm and 10-20mm are the most numerous, constituting a significant portion at 40.33\% and 39.78\% of all nodules, individually. In contrast, pulmonary nodules within the 20-30 mm and >30 mm size ranges represent relative smaller proportions, accounting for only 14.37\% and 5.52\%, respectively, as visually depicted in Fig. \ref{fig:Fig. 5}(b). 

Fig. \ref{fig:Fig. 5}(c) illustrates the reciprocal correlation between nodule size and pathologic type. Notably, it becomes evident that invasive adenocarcinomas dominate the category of nodules with sizes ranging from 20 to 30 mm.

\textbf{Thickness:} We have concurrently collected CT sequences with slice thicknesses of 1.25mm and 5mm to address diverse requirements in actual clinical diagnosis and treatment.

\subsection*{Data Preprocessing}

Initially, Gaussian filtering is applied to each image slice to achieve two objectives: image smoothing for noise reduction and enhancement of color contrast between the lungs and surrounding tissues. Subsequently, a clustering technique is employed to determine thresholds that effectively distinguish lung regions from non-lung regions within the medical images. The ultimate aim of this preprocessing step is to isolate and highlight the lung region, so that the structure or region of interest can be more easily identified in successive segmentation and analysis steps. The processed lung region is then systematically stored in a designated folder, labeled as “lung\_mask”.

\section*{Data Records}

We have made the data available in a public repository without any password protection. Due to the larger size of our dataset, it has been divided and stored on two separate web pages for ease of access and efficient handling. Meanwhile, For the sake of facilitating downloads, particularly large files will be divided into multiple compressed archives. The dataset can be accessible through the following links, Database Part A: https://zenodo.org/record/8354947; Database Part B: https://zenodo.org/record/8357533. 

For the layer thickness of 1.25 mm, two different types of datasets are provided for classification and detection. The detailed description about the file format is given below:

\textbf{1.25mm slice thickness classification dataset:} The classification dataset, comprising BMP-format images, has been compressed into a single archive named “class\_dataset.zip” and subsequently uploaded to Database Part A. Within this compressed archive, the dataset has been partitioned into two distinct segments, the training set and the testing set, in a 5:1 ratio. These segments are respectively stored in the “train” and “validate” folders. Each folder contains six subfolders that are categorized into label 1, 2, 3, 4, 5 and 6, correspondingly. These six distinct categories correspond to invasive adenocarcinoma, minimally invasive adenocarcinoma, adenocarcinoma in situ, malignant other subtypes, inflammation, and benign other subtypes, respectively. In addition, the “READ\_ME.txt” file provides detailed information including the number of samples in the training and testing sets for different categories.

\textbf{1.25mm slice thickness detection dataset: }Regarding 1.25mm slice thickness, this study has also developed a dataset for lung nodule detection and evaluation, encompassing both three-dimensional (3D) and two-dimensional (2D) data formats. 

In the context of 3D data, we have provided two distinct data formats: MHD and BMP. In specific terms, the dataset based on the MHD format comprises four compressed files, which have been uploaded to Database Part A. Among these, “5mm\_3D\_detection\_mhd0.zip” and “5mm\_3D\_detection\_mhd1.zip” contain the original MHD-format images of the patients, “5mm\_3D\_detection\_mhd2.zip” comprises MHD-format images that have undergone lung segmentation and “5mm\_3D\_detection\_mhd3.zip” contains “annotations.csv” that holds information on the actual nodule positions and diameters. These data points are referenced within a global coordinate system, serving as a reliable benchmark for assessing the performance of the testing data. 

Conversely, the detection dataset in BMP format comprises data that has undergone augmentation through flipping and rotation, and it is stored in dataset part B. We have compressed the images into three distinct sub-archives for easy upload, each labelled as “1.25mm\_3D\_detection\_bmp\_imgs0.zip”, “1.25mm\_3D\_detection\_bmp\_imgs1.zip” and “1.25mm\_3D\_detection\_bmp\newline\_imgs2.zip”. Researchers should extract and consolidate these compressed files for utilization. Furthermore, within the “1.25mm\_3D\_detection\_bmp3.zip”, there exists a file named “image\_annotation\_augment\_diameter.csv”. This file provides detailed information regarding the real nodule positions and diameters based on the voxel coordinate system. The data from “image\_annotation\_augment\_diameter.csv” has been divided into training and testing sets in a 5:1 ratio. They are respectively saved as “train\_anno.csv” and “val\_anno.csv” files, and the case numbers for the training and testing sets have been written into “train.txt” and “val.txt”, independently.

With respect to the 2D data with the layer thickness of 1.25mm, we provide the “1.25mm\_2D\_detection.zip” folder in Database Part A, comprising the “JPEGImages” folder housing the dataset’s images and the “Annotations” folder containing XML-format annotations. Furthermore, we have presented image filenames with nodule coordinates (xmin, ymin, xmax, ymax) in “VOC\_CT.xlsx”. To facilitate model training and performance evaluation, we partitioned the dataset into training and testing sets using a 4:1 ratio. The filenames of these sets have been recorded in the “train.txt” and “val.txt” files located within the “ImageSets/Main” directory, while the image paths are documented in “train.txt” and “val.txt” within the “images” folder. Furthermore, we have included the image paths and corresponding nodule coordinates in the “train.txt” and “val.txt” files inside the “1.25mm\_2D\_detection.zip”. 

\textbf{5mm slice thickness detection dataset:} For CT sequences with a slice thickness of 5mm, we have also furnished a detection dataset availablely in MHD format within Database Part A. The dataset has been compressed into four files. The first three files, namely “5mm\_3D\_detection\_mhd0 (or 1, 2).zip”, contain data in MHD format. The final file, “5mm\_3D\_detection\_mhd3.zip”, provides information about the actual tuberculosis locations and diameters for each case in the global coordinate system.

\section*{Technical Validation}

\subsection*{Classification experiments}

To appraise the efficacy of our designed pulmonary nodule classification dataset, we conduct comprehensive assessments utilizing eight distinct and typical image classification networks, namely ResNet \cite{he2016deep}, ConvNext \cite{liu2022convnet}, ResNext \cite{xie2017aggregated}, Res2Net \cite{gao2019res2net}, SE-ResNet \cite{hu2018squeeze}, CABNet \cite{he2020cabnet}, InceptionV4 \cite{szegedy2017inception}, and EfficientNet \cite{tan2019efficientnet}. These networks exhibit diverse architectures and excellent performance characteristics, enabling a thorough examination of the dataset’s suitability for pulmonary nodule classification tasks. We carry out both training and testing phases for these networks using identical datasets and configurations to ensure the fairness and comparability of our experiments. Each dataset is divided into training and testing sets in a 5:1 ratio. During the experiment, each network undergoes training for a total of 150 epochs.

For the quantitative assessment of classification performance, we employ two key metrics, namely Extreme Accuracy (ACC) and Quadratic Weighted Kappa (QWK), for objective evaluation. These metrics are selected to provide a thorough assessment of our classification results. The outcomes of these experiments are presented in Table \ref{tab:table 1}.

\begin{table}[ht]
\centering
\begin{tabular}{|l|l|l|l|l|l|l|l|l|}
\hline
& ResNet\cite{he2016deep} & ConvNext\cite{liu2022convnet}
 & ResNext\cite{xie2017aggregated}
 & Res2Net\cite{gao2019res2net}
 & SEResNet\cite{hu2018squeeze}
 & CABNet\cite{he2020cabnet}
 & InceptionV4\cite{szegedy2017inception}
 & EfficientNet\cite{tan2019efficientnet}
\\
\hline
ACC  & 0.6103 & 0.5787 & 0.6054 & 0.5441 & 0.5613 & 0.5711 & 0.4436 & 0.5441\\
\hline
QWK  & 0.5413  & 0.4662 & 0.5525 & 0.5238 & 0.5553 & 0.3717 & 0.3368 & 0.4149\\

\hline
\end{tabular}
\caption{\label{tab:table 1}Performance assessment based on eight representative classification models.}
\end{table}

As evident from the table, our dataset yields reliable results across various categorical networks. This consistency underscores the dataset’s label and feature accuracy, as well as its robustness. The multi-model applicability indicates that the dataset encompasses diverse data samples and has been adequately considered in terms of data quality and diversity. While all eight typical networks achieve accuracy and QWK scores exceeding 0.5, it is remarkably that none of their metrics attain exceptionally high values. This observation accentuates the challenges and research potential inherent in our dataset.

\begin{figure}[ht]
\centering
\includegraphics[width=\linewidth]{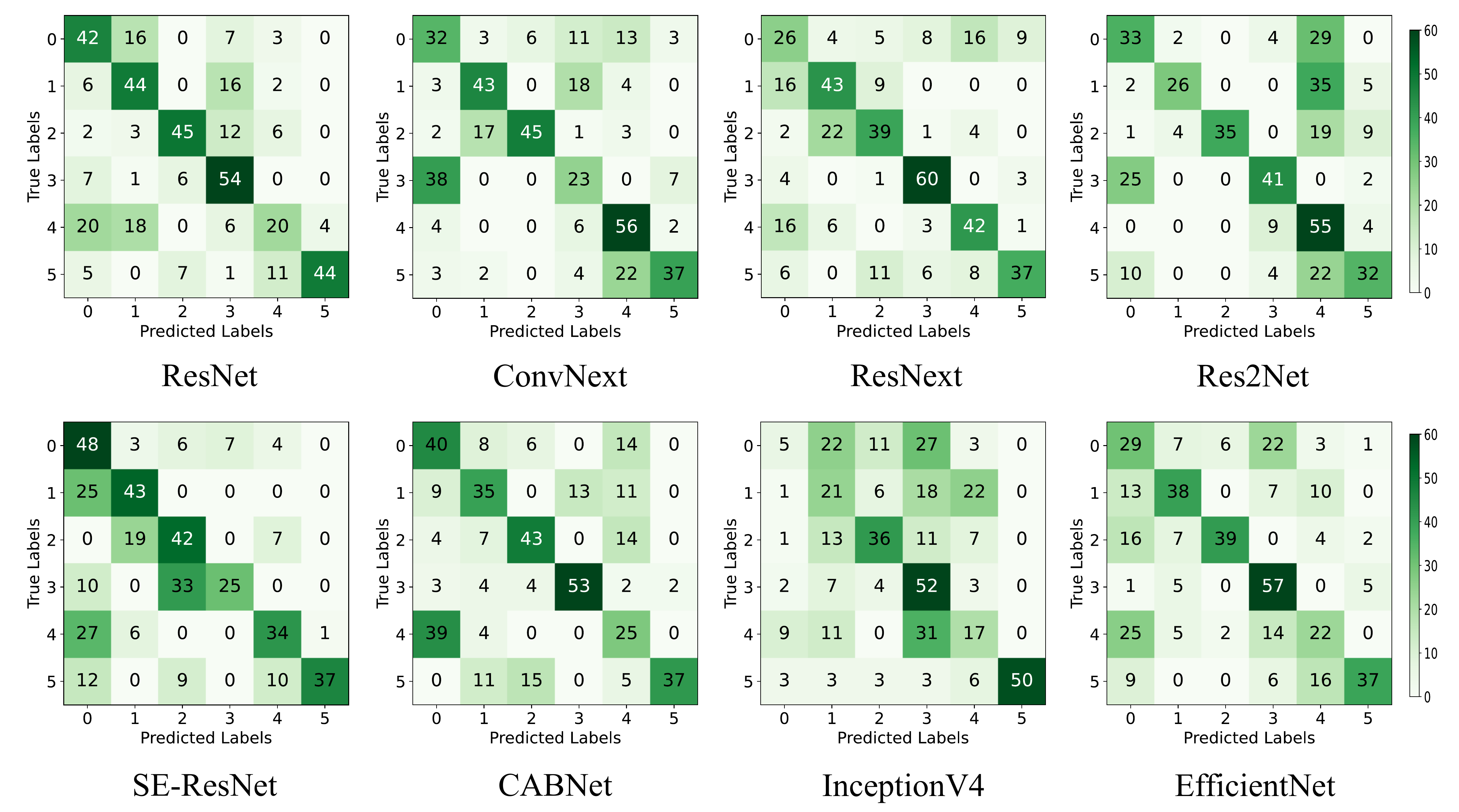}
\caption{The confusion matrix between the real label and the predicted result.
}
\label{fig:Fig. 6}
\end{figure}

To assess the quality of category detection for the eight individual networks, we present the confusion matrices in Fig. \ref{fig:Fig. 6}. Significantly, the values along the main diagonal of each confusion matrix are relatively higher, signifying that the network classification results generally aligned with our annotations. This alignment highlights the effectiveness of our dataset for classification tasks. Given that different networks exhibit varying sensitivities to distinct data classes, this further verifies the dataset’s complexity and diversity.

\subsection*{Detection experiments}

In order to validate the effectiveness of the lung nodule detection dataset, we test five different object detection networks, consisting of Faster R-CNN \cite{ren2015faster}, Yolov3 \cite{redmon2018yolov3}, MobileNet \cite{howard2017mobilenets}, SSD \cite{liu2016ssd}, and RetinaNet \cite{lin2017focal}. Remarkably, each network is trained over 50 epochs using the same dataset and train configuration. To estimate the model’s performance, we employ object detection evaluation metrics typically used for the COCO dataset, including Average Precision (AP), $AP_{50}^{}$, $AP_{75}^{}$, $AP_{S}^{}$, $AP_{M}^{}$, and $AP_{L}^{}$ \cite{redmon2018yolov3}. These metrics are widely recognized for assessing the accuracy of detection box localization by the model. Detailed test results are presented in Table \ref{tab:table 2}.

\begin{table}[ht]
\centering
\begin{tabular}{|l|l|l|l|l|l|l|}
\hline
& AP & $AP_{50}^{}$
 & $AP_{75}^{}$
 & $AP_{S}^{}$
 & $AP_{M}^{}$
 & $AP_{L}^{}$
\\
\hline
Faster R-CNN \cite{ren2015faster} & 0.772 & 0.969 & 0.840 & 0.733 & 0.815 & 0.854\\
\hline
Yolov3 \cite{redmon2018yolov3} & 0.669  & 0.958 & 0.765 & 0.574 & 0.764 & 0.828\\
\hline
MobileNet \cite{howard2017mobilenets} & 0.645 & 0.954 & 0.731 & 0.562 & 0.738 & 0.823 \\
\hline
SSD \cite{liu2016ssd} & 0.757 & 0.980 & 0.862 & 0.712 & 0.809 & 0.871 \\
\hline
RetinaNet \cite{lin2017focal} & 0.810 & 0.980 & 0.876 & 0.769 & 0.858 & 0.934 \\
\hline
\end{tabular}
\caption{\label{tab:table 2}Evaluation results in terms of different detection models.}
\end{table}

Specifically, $AP_{S}^{}$, $AP_{M}^{}$, and $AP_{L}^{}$ gauge the detection performance of target objects at diverse sizes. Given that our dataset contains over twice as many malignant nodules as benign nodules, the substantial number of malignant nodules enhances the model’s sensitivity in detecting malignancies. Consequently, the $AP_{L}^{}$ metrics for the results obtained with the five popular networks are higher.

\section*{Code availability}

To make this dataset easily available to users, we have released a resource located in a GitHub repository. The repository contains Python sample code for data type conversion and data manipulation to help users better understand and utilise this dataset. You can find the resource in the following GitHub repository: https://github.com/Heroandzhang/dataset\_process.

\section*{Competing interests} 

The authors declare that they have no competing interests.

\section*{Research involving Human Participants and/or animals}

The authors declare that they have no involving Human Participants and/or Animals to this work. 

\bibliography{sample}

\end{document}